\documentclass[twocolumn,aps,prb,final,amsfonts,amssymb,amsmath,%
floatfix,showpacs,superscriptaddress,byrevtex]{revtex4}

\usepackage[]{graphicx}
\usepackage{bm}

\newcommand{\etal}{\textit{et al.}}
\newcommand{\eg}{e.g., }

\begin{document}

\title{Nuclear spin effects in singly negatively charged InP quantum dots}

\author{Bipul~Pal}
\email[E-mail: ]{bipulpal@sakura.cc.tsukuba.ac.jp}
\affiliation{Institute of Physics and Tsukuba Advanced Research
Alliance (TARA), University of Tsukuba, Tsukuba 305-8571, Japan}

\author{Sergey~Yu.~Verbin}
\email[E-mail: ]{syuv54@mail.ru} \affiliation{Institute of
Physics, St.-Petersburg State University, St.-Petersburg, 198504,
Russia} \affiliation{Venture Business Laboratory, University of
Tsukuba, Tsukuba 305-8571, Japan}

\author{Ivan~V.~Ignatiev}
\affiliation{Institute of Physics, St.-Petersburg State
University, St.-Petersburg, 198504, Russia} \affiliation{Venture
Business Laboratory, University of Tsukuba, Tsukuba 305-8571,
Japan}

\author{Michio~Ikezawa}
\affiliation{Institute of Physics and Tsukuba Advanced Research
Alliance (TARA), University of Tsukuba, Tsukuba 305-8571, Japan}

\author{Yasuaki~Masumoto}
\affiliation{Institute of Physics and Tsukuba Advanced Research
Alliance (TARA), University of Tsukuba, Tsukuba 305-8571, Japan}
\date{\today}

\begin{abstract}
Experimental investigation of nuclear spin effects on the electron
spin polarization in singly negatively charged InP quantum dots is
reported. Pump-probe photoluminescence measurements of electron
spin relaxation in the microsecond timescale are used to estimate
the time-period $T_{N}$ of the Larmor precession of nuclear spins
in the hyperfine field of electrons. We find $T_{N}$ to be $\sim
1$~$\mu$s at $T \approx 5$~K, under the vanishing external
magnetic field. From the time-integrated measurements of electron
spin polarization as a function of a longitudinally applied
magnetic field at $T \approx 5$~K, we find that the Overhauser
field appearing due to the dynamic nuclear polarization increases
linearly with the excitation power, though its magnitude remains
smaller than $10$~mT up to the highest excitation power ($50$~mW)
used in these experiments. The effective magnetic field of the
frozen fluctuations of nuclear spins is found to be $15$~mT,
independent of the excitation power.
\end{abstract}


\maketitle

\section{Introduction} \label{intro}
Strong localization of electrons in quantum dots (QDs) may enhance
the hyperfine interaction of electron spins with those of
nuclei.~\cite{gammonprl86} Various aspects of the hyperfine
interaction of electron and nuclear spins have been studied for
last three decades in different materials,~\cite{optorientbook}
including InP QDs.~\cite{dzhioevjetpl68,dzhioevpss41}
Charge-tunable InP QDs with one resident electron per dot, on an
average, have recently attracted considerable research interest
due to the observation of submillisecond spin lifetime of resident
electrons in these QDs.~\cite{ikezawaprb72,paljpsj75} This
observation makes it a promising candidate for quantum memory
element in the emerging fields of quantum information technology
and spintronics.~\cite{spintronicsbook} Application of QDs as the
building blocks of quantum computers has been
proposed.~\cite{losspra57}

However, the influence of the hyperfine interaction between
electron and nuclear spins, on the long-lived electron spin
polarization needs to be clarified. Two effects of the
electron-nuclear spin-spin interactions may be considered. One of
them is the so-called dynamic nuclear
polarization.~\cite{optorientbook} In the optical orientation of
electron spins in semiconductors by circularly polarized photons,
the spin-oriented electrons dynamically polarize the nuclear spins
due to the hyperfine coupling of the electron and nuclear spin
subsystems.~\cite{optorientbook,berkovitsspjetp38} In turn, the
spin-polarized nuclei produce an effective internal magnetic field
(the Overhauser field $B_{N}$), which may influence the electron
spin dynamics. In presence of an external magnetic field
$B_{\text{ext}}$, electron spins should feel a total magnetic
field $B_{T}=B_{\text{ext}}+B_{N}$.

Another important effect of the electron-nuclear hyperfine
coupling is the electron spin relaxation via its interaction with
nuclear spins.~\cite{merkulovprb65,khaetskiiprl88,semenovprb67} At
low temperature, relaxation of the coupled electron-nuclear spin
system is determined by three processes, namely, i) electron spin
precession in the frozen fluctuations of the hyperfine field of
the nuclear spins, ii) nuclear spin precession in the hyperfine
field of the electron spins, and iii) nuclear spin relaxation in
the dipole-dipole field of its nuclear neighbors. These three
processes have very different characteristic timescales.
Theoretical estimate for GaAs QDs containing $\sim 10^{5}$ nuclei
suggest them to be $\sim 1$~ns, $\sim 1$~$\mu$s, and $\sim
100$~$\mu$s, respectively.~\cite{merkulovprb65} The last one of
the above processes will not be considered further in this paper,
because it affects the electron spin dynamics over a long
timescale of $10^{-4}$--$10^{-3}$~s. In this long time-regime many
other mechanisms, such as those originating from the spin-orbit
coupling and interaction with phonons become important for
electron spin relaxation.

The first relaxation process mentioned above arises from the fact
that due to a large, but limited number of nuclear spins,
typically $n\sim10^5$, interacting with the electron spin in a QD,
random correlation of nuclear spins may create a fluctuating
nuclear polarization, $\Delta F_{N} \propto F_{N}/\sqrt{n}$, where
$F_{N}$ is the total spin of the polarized nuclei. Fluctuation
$\Delta F_{N}$ acts on the electron spin as another internal
magnetic field $B_{f}$, with random magnitude and orientation over
the QD ensemble.~\cite{merkulovprb65} We may note that the Larmor
precession of the nuclear spins in the hyperfine field of an
electron spin is much slower than that of the electron spin in the
nuclear hyperfine field, because the interaction of an electron
spin with a single nucleus is $\sqrt{n}$ times weaker compared to
its interaction with the effective magnetic field of the nuclear
fluctuations. Thus, the electron ``sees'' a snapshot of the
``frozen fluctuations'' of the nuclear field. Electron spin
precession in the magnetic field $B_{f}$ is expected to cause
electron spin relaxation in the QD ensemble in a timescale of the
order of 1~ns, during which the electron spin polarization decays
to one-third of its initial value. After the initial decay in the
nanosecond timescale, the remaining spin polarization relaxes very
slowly over a timescale much longer than the radiative lifetime of
the electron-hole pair.~\cite{merkulovprb65}

Electron spin relaxation due to the frozen fluctuations of nuclear
spins (FFNS) may be suppressed by a longitudinally (along the
optical excitation axis) applied magnetic field with a magnitude
larger than $B_{f}$.~\cite{merkulovprb65} Recent study of electron
spin relaxation in $p$-doped InAs QDs by Braun
{\etal}~\cite{braunprl94} found that at zero external magnetic
field, electron spin polarization decays down to one-third of its
initial value within 800~ps after photo-excitation. Then the
residual spin polarization remains stable with no measurable decay
within the photoluminescence (PL) decay time. The authors found
that the initial fast relaxation of electron spin was suppressed
by the application of a small ($\sim 100$~mT) external magnetic
field. We want to note the fact that for $B_{\text{ext}} = 0$,
{\emph{two distinctly different time regimes are present}} in the
electron spin relaxation.~\cite{braunprl94fig3} The initial fast
relaxation is caused by the random distribution of $B_{f}$ in the
QD ensemble.~\cite{merkulovprb65,braunprl94} However, {\emph{a
total depolarization of electron spins does not take place}} in
this regime. The residual electron spin polarization decays very
slowly in a second relaxation regime, which is governed by the
slow time-varying change in the distribution of $B_{f}$. This slow
change in the distribution of $B_{f}$ is caused by the variation
in the precession rate of nuclear spins in the hyperfine field of
electrons.~\cite{merkulovprb65,braunprl94} We denote it as the
nuclear spin precession effect (NSPE). Study of electron spin
relaxation due to the NSPE in the microsecond time-range should
give an estimate of the nuclear spin precession period $T_{N}$ in
the hyperfine field of electrons. However, to the best of our
knowledge, this has not been experimentally studied so far.

In this paper we describe our experimental study of nuclear spin
effects on the long-lived spin polarization of resident electrons,
observed recently~\cite{ikezawaprb72,paljpsj75} in the singly
negatively charged InP QDs. Electron spin dynamics and the
influence of nuclear spins on it, are probed by the time-resolved
as well as time-integrated measurements of the degree of PL
circular polarization $\rho_{c}$, defined quantitatively in
Sec.~\ref{expt}. Our time-domain measurements of $\rho_{c}$ by
using a PL pump-probe
technique~\cite{ikezawaprb72,paljpsj75,cortezprl89,coltonprb69} in
the microsecond time-regime reveal electron spin relaxation via
the NSPE at the vanishing external magnetic field. From the value
of the electron spin decay time $\tau_{d}$ for $B_{\text{ext}} =
0$, we estimate that $T_{N} \sim 1$~$\mu$s at $T \approx 5$~K,
comparable to the theoretical estimate available for GaAs
QDs.~\cite{merkulovprb65} We also measure the dependence of
$\rho_{c}$ on $B_{\text{ext}}$ in time-integrated measurements at
$T \approx 5$~K. From these steady-state measurements we obtain
the Overhauser field $B_{N} = 6$~mT at 50~mW CW excitation and the
effective magnetic field of the FFNS, $B_{f} = 15$~mT, independent
of the excitation power. The relatively small value of $B_{N}$ may
come from efficient leakage of QD nuclear spin polarization to the
surrounding lattice nuclei.

\section{Experimental} \label{expt}
Our sample consists of a single layer of self-assembled InP QDs,
embedded between Ga$_{0.5}$In$_{0.5}$P barriers grown on a
$n^{+}$-GaAs substrate. The average base diameter (height) of the
QDs is $\sim 40$ (5)~nm and the areal density of dots is $\sim
10^{10}$~cm$^{-2}$. Semi-transparent indium-tin-oxide electrode is
deposited on top of the sample to control the charge state of the
QDs by means of an applied electric
bias.~\cite{ikezawaprb72,paljpsj75,kozinprb65} For the present
study on the singly negatively charged QDs we apply an electric
bias of $U_{b} = -0.1$~V. This is because it was found from a
previous study of trionic quantum beats~\cite{kozinprb65} on the
same sample that at $U_{b} \approx -0.1$~V the QDs contain one
resident electron per dot, on an average.

Electron spins in the QD ensemble are polarized in our experiments
by using the well-known optical orientation
technique.~\cite{optorientbook,paljpsj75model} We excite the QDs
{\emph{quasiresonantly}} (in the excited state of the QDs, but
below the wetting layer bandgap) by circularly polarized beam from
a Ti:Sapphire laser, which can be operated either in CW or in
pulsed mode. The excitation beam is directed along the sample
growth axis and is focused to a spotsize of $\sim 150$~$\mu$m in
diameter on the sample kept in a magneto-optical cryostat at
$T\approx 5$~K. The excitation energy $E_{x} = 1.77$~eV and the
detection energy $E_{d} = 1.72$~eV used in our experiments are
indicated by arrows on the polarization-resolved PL spectra in
Fig.~\ref{plspec}(inset). These spectra are measured by using
suitable combinations of retardation plates and Glan-Thompson
linear polarizers and a triple spectrometer (focal length $1$~m)
equipped with a CCD detector. The spectral resolution is
$0.05$~meV. We monitor the degree of circular polarization
$\rho_{c}=(I_{S}-I_{O})/(I_{S}+I_{O})$ for the ground state PL.
Here $I_{S}$~($I_{O}$) is the intensity of the PL component having
the same (opposite) circular polarization as that of the
excitation beam. We study $\rho_{c}$, in the time-integrated as
well as time-resolved measurements, as a function of the external
magnetic field $B_{\text{ext}}$ applied along the optical
excitation axis (longitudinal magnetic field, Faraday geometry).
Time-resolved data are taken by using a synchroscan streak camera,
while for the time-integrated measurements, a GaAs photomultiplier
tube and a two channel gated photon counter are used.

\section{Results and discussion}
\subsection{Negative circularly polarized PL}
\begin{figure}[tbh]
\includegraphics[clip,width=7.5cm]{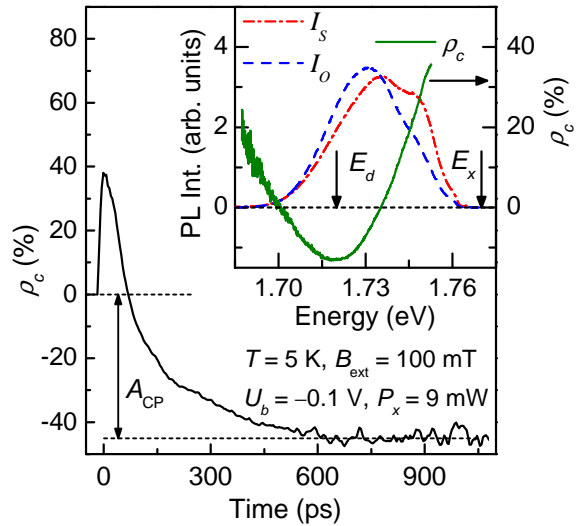}
\caption{\label{plspec}(Color online) Time-dependence of
$\rho_{c}$ measured for $E_{x}$ and $E_{d}$ indicated in the
inset. $B_{\text{ext}} = 100$~mT was applied to suppress the
effect of the FFNS. The inset shows the spectra of $I_{S}$ and
$I_{O}$. Spectral dependence of $\rho_{c}$ is shown for $E_{x}$
indicated by an arrow.}
\end{figure}

Our measurements of polarization-resolved PL spectra under
quasiresonant excitation ($E_{x} = 1.77$~eV) of singly negatively
charged InP QDs show that the degree of circular polarization
$\rho_{c}$ is
{\emph{negative}}~\cite{ikezawaprb72,paljpsj75,masumotoprb74,others}
in the spectral region $1.7$--$1.735$~eV, for which $\Delta E =
(E_{x} - E_{d})$ lies between $70$--$35$~meV
[Fig.~\ref{plspec}(inset)]. We measure the time dependence of
$\rho_{c}$ for $E_{d} = 1.72$~eV. The data is plotted in
Fig.~\ref{plspec}. Initially $\rho_{c}$ is seen to be positive,
but it becomes negative at $70$~ps after the excitation pulse and
then $\rho_{c}$ approaches a {\emph{constant negative value}}. We
denote this constant value as the amplitude of circular
polarization of PL ($A_{\text{CP}}$) [see Fig.~\ref{plspec}]. The
value of $A_{\text{CP}}$ increases logarithmically with the
excitation power~\cite{optpump} and reaches up to $45${\%}. In the
time-integrated measurements negative value of $\rho_{c}$ reaches
up to $30${\%}.

\begin{figure}[tbh]
\includegraphics[clip,width=8.0cm]{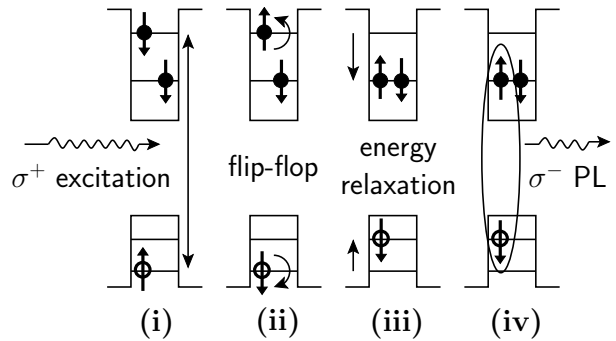}
\caption{\label{ncpmodel} A schematic model to explain the
occurrence of the negative value of $\rho_{c}$ due to the presence
of the optically oriented resident electron spins in the QDs.}
\end{figure}

The remarkable {\emph{negative circular polarization of PL}} is
related to the optically created spin orientation of the resident
electrons. How the sign of $\rho_{c}$ is determined by the spin
direction of the resident electron in a QD is discussed in many
papers, see {\eg} Refs.~\onlinecite{ikezawaprb72,paljpsj75,%
cortezprl89,laurentpe20,wareprl95,kavokinpssa195}. Here we briefly
explain how the negative degree of circular polarization of PL
arises for our experimental condition, due to the presence of the
optically polarized resident electron spins in the QDs. For this,
we refer to the schematic diagram of Fig.~\ref{ncpmodel}.
Quasiresonant excitation in our experiments creates electron-hole
pair in the QD excited state. As the resident electron spins in
the QDs are polarized by the excitation
photons,~\cite{paljpsj75model} the spin of the photogenerated
electron in the excited state and that of the resident electron in
the ground state should have a parallel orientation in the
majority of the QDs in the ensemble. For simplicity, we consider
here only these QDs [Fig.~\ref{ncpmodel}(i)], as $\rho_{c}$ for
the QD ensemble is mainly determined by them. In these QDs, a
direct energy relaxation of the photogenerated electron to the
ground state is blocked due to Pauli exclusion principle. However,
the electron-hole pair in the excited state can undergo a
flip-flop transition,~\cite{paljpsj75,laurentpe20,kalevichprb72}
in which a simultaneous flip of the electron and hole spins takes
place [Fig.~\ref{ncpmodel}(ii)]. This is followed by energy
relaxation of both the hot carriers [Fig.~\ref{ncpmodel}(iii)].
The flip-flop transition is caused by the anisotropic exchange
interaction in QDs.~\cite{laurentpe20,wareprl95,kavokinpssa195}
After the flip-flop transition, PL emitted from the radiative
recombination of the spin-flipped electron and hole in the ground
state has the opposite circular polarization compared to the
excitation photons [Fig.~\ref{ncpmodel}(iv)]. Thus, the degree of
PL circular polarization becomes negative. We assume that the hole
spin relaxation time in the QD ground state is much longer than
the radiative lifetime.~\cite{flissikowskiprb68,laurentprl94} This
is supported by the data of Fig.~\ref{plspec}, where $\rho_{c}$
approaches a constant negative value and remains stable over the
PL decay time.

The measurements of $\rho_{c}$ in the time-integrated experiments
and of $A_{\text{CP}}$ in the time-resolved experiments are used
in this work for the study of nuclear spin effects on the spin
polarization of resident electrons in the singly negatively
charged QDs.

\subsection{Frozen fluctuations of nuclear hyperfine field}
\begin{figure}[tbh]
\includegraphics[clip,width=6.8cm]{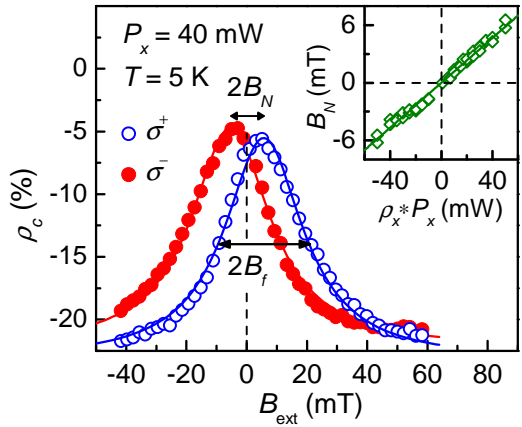}
\caption{\label{dnp}(Color online) Dependence of $\rho_{c}$ on
$B_{\text{ext}}$ for $\sigma^{+}$- and $\sigma^{-}$-polarized CW
excitations. The solid lines are Lorentzian fits used to estimate
$B_{f}$. Shifts of the minima of $\left| \rho_{c} \right|$ from
$B_{\text{ext}}=0$ estimate $B_{N}$. The inset shows linear
dependence of $B_{N}$ on the scaled excitation power
$\rho_{x}\!*\!P_{x}$, where $\rho_{x} = +1$~($-1$) for
$\sigma^{+}$~($\sigma^{-}$) excitation and $P_{x}=$~excitation
power.}
\end{figure}

It turns out that the PL circular polarization, and hence, the
electron spin polarization, is very sensitive to $B_{\text{ext}}$.
Time-integrated measurements in Fig.~\ref{dnp} show the dependence
of $\rho_{c}$ on $B_{\text{ext}}$ for $\sigma^{+}$- and
$\sigma^{-}$-polarized CW excitations. As seen there, $\rho_{c}$
becomes nearly independent of $B_{\text{ext}}$ for $B_{\text{ext}}
> 50$~mT. Absolute value of $\rho_{c}$ decreases with decreasing $\left|
B_{\text{ext}} \right|$ and reaches a minimum for $\left|
B_{\text{ext}} \right|$ nearly, but not exactly zero. The behavior
of $\rho_{c}$ as a function of $B_{\text{ext}}$ can be fitted well
by a Lorentzian with a half width at half maximum of $15$~mT.

The decrease of $\left| \rho_{c} \right|$ may be interpreted as
the effect of electron spin relaxation in the QD ensemble by the
field $B_{f}$ of the FFNS.~\cite{merkulovprb65} The effect is
suppressed by the external magnetic field when $B_{\text{ext}}$
exceeds $B_{f}$ in magnitude, allowing electron spin polarization,
and hence, $\left| \rho_{c} \right|$ to increase to reach a
steady-state value. So, the value of $B_{f}$ may be estimated from
the half width at half maximum of the Lorentzians in
Fig.~\ref{dnp}. Thus, we estimate a value of $B_{f} = 15$~mT for
the QDs under study. The obtained value of $B_{f}$ is found to be
independent of the excitation power up to the highest excitation
power ($50$~mW) used, suggesting that it is intrinsic to the InP
QDs. For $B_{f} = 15$~mT, we estimate the electron spin relaxation
time $\tau_{s} = \hbar / (g_{e} \mu_{B} B_{f}) \approx 0.5$~ns
($\mu_{B} =$~Bohr magneton and $g_{e} = \text{electron Land{\'e}
g-factor} = 1.5$~\cite{masumotojlum108}), resulting from the FFNS.
The values of $B_{f} = 15$~mT and $\tau_{s} = 0.5$~ns obtained
here for the InP QDs, are comparable to those estimated
theoretically by Merkulov {\etal}~\cite{merkulovprb65} for GaAs
QDs, and to those obtained experimentally by Braun
{\etal}~\cite{braunprl94} for InAs QDs.

\subsection{Dynamic nuclear polarization}
As seen in Fig.~\ref{dnp}, the minima of $\left| \rho_{c} \right|$
is shifted from $B_{\text{ext}}=0$. This is caused by the
Overhauser field $B_{N}$, due to which the electron spin ``feels''
a total magnetic field $B_{T}=B_{N}+B_{\text{ext}}$ in presence of
the external magnetic field $B_{\text{ext}}$. When averaged over
the QD ensemble, the field $B_{f}$ of the FFNS does not contribute
to the total magnetic field. However, it causes electron spin
relaxation, and hence, decrease in $\left| \rho_{c} \right|$ when
$B_{T}$ approaches zero. Thus, the minimum of $\left| \rho_{c}
\right|$ should be observed at $B_{\text{ext}}=-B_{N}$. This
allows us to estimate $B_{N}$. The sign of $B_{N}$ created by
light should be opposite for the $\sigma^{+}$- and
$\sigma^{-}$-polarized excitations. As a result, the minima of
$\left| \rho_{c} \right|$ for $\sigma^{+}$- and
$\sigma^{-}$-polarized excitations are shifted symmetrically from
$B_{\text{ext}}=0$ in opposite directions in
Fig.~\ref{dnp}.~\cite{altexct} The minima in the two cases differ
by $2B_{N}$. We study the excitation power ($P_{x}$) dependence of
$B_{N}$. A plot of $B_{N}$ as a function of the scaled excitation
power $\rho_{x}\!*\!P_{x}$, where the helicity $\rho_{x} =
+1$~($-1$) for $\sigma^{+}$~($\sigma^{-}$) excitation, shows that
the dynamic nuclear polarization builds up linearly with the
excitation laser power [Fig.~\ref{dnp}(inset)].

We find experimentally that up to $P_{x} = 50$~mW, $B_{N}$ remains
smaller than $10$~mT, and that $B_{N}$ does not show any
indication of saturation. The value of $B_{N} = 6$~mT observed at
$P_{x} = 50$~mW in our experiments is in agreement with previous
reports of $B_{N}$ measured in an ensemble of InP nano-islands
embedded in InGaP matrix.~\cite{dzhioevjetpl68} A value of $B_{N}
= 1.2$~T for GaAs QDs formed by interface nano-roughness in a GaAs
quantum well has been reported by Gammon
{\etal}~\cite{gammonprl86} and Brown {\etal}~\cite{brownprb54} in
single dot measurements, where they estimated that $65${\%} of the
nuclear spins were polarized. Yokoi {\etal}~\cite{yokoiprb71} have
reported $B_{N} = 160$~mT by using single dot spectroscopy for
self-assembled InAlAs QDs, where $6${\%} of the nuclei were
polarized. Origin of the small $B_{N}$ observed for self-assembled
InP QD ensemble is not fully clear. We may say that only a small
fraction of the nuclei in a QD are polarized, because we do not
observe any saturation of $B_{N}$ up to the highest excitation
power used. This may be caused by the low excitation efficiency of
the QDs under quasiresonant excitation we used and also by the
inefficient transfer of spin polarization from electrons to
nuclei.~\cite{imamogluprl91} Another possible reason may be the
efficient leakage of nuclear spin polarization from the QDs to the
surrounding lattice nuclei.~\cite{optorientbook} Due to the large
nuclear spin of In ($I=9/2$), it may loose its spin polarization
rather efficiently through quadrupole interaction in presence of a
time varying gradient of local electric field, which may be
created by the photogenerated electrons in the
QDs.~\cite{salisprb64} Then, due to close proximity, spin
polarization of P nuclei ($I=1/2$) may be transferred to In nuclei
and eventually lost to the surrounding lattice nuclei due to
efficient spin relaxation of In nuclei.

\subsection{Electron spin relaxation by slow variation in
frozen fluctuations of nuclear spins} 
\begin{figure}[tbh]
\includegraphics[clip,width=7.0cm]{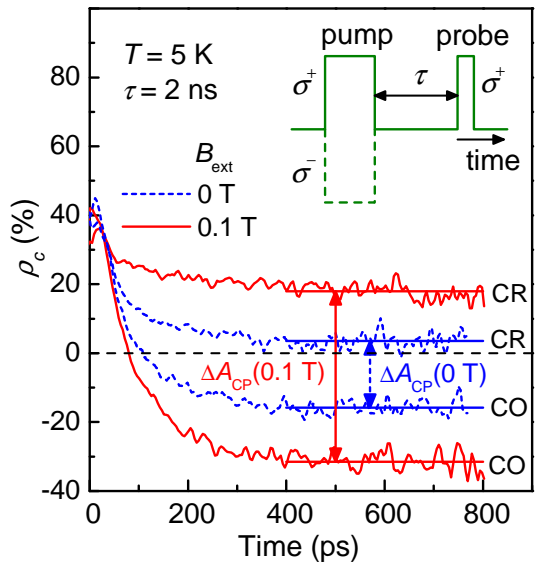}
\caption{\label{trpl}(Color online) Kinetics of $\rho_{c}$ for the
probe PL at $\tau = 2$~ns for $B_{\text{ext}}=0$ and $0.1$~T. The
pump and probe beams are either co-circularly (CO) or
cross-circularly (CR) polarized. The difference $\Delta
A_{\text{CP}}$ between $A_{\text{CP}}$ for the CR- and CO-cases
are indicated by arrows for $B_{\text{ext}}=0$ and $0.1$~T. It is
found that $\Delta A_{\text{CP}} \text{(0~T)} / \Delta
A_{\text{CP}} \text{(0.1~T)} \approx 1/3$. The inset shows a
schematic of the pump and probe pulse configurations used in the
polarization-resolved pump-probe PL experiment.}
\end{figure}

For a direct time-domain
study of electron spin relaxation we use a pump-probe PL
technique.~\cite{ikezawaprb72,paljpsj75,cortezprl89,coltonprb69}
One of the main advantage of this technique is that the measurable
time-range of the spin dynamics by this method is not limited by
the PL lifetime, in contrast to the time-resolved PL measurements
where spin relaxation can be monitored only within the PL decay
time.~\cite{braunprl94} Details of our pump-probe PL experimental
arrangement are discussed in
Refs.~\onlinecite{ikezawaprb72,paljpsj75}. In this method, a
circularly polarized ($\sigma^+$ or $\sigma^-$) pump pulse creates
a spin orientation of the resident
electrons.~\cite{paljpsj75model} The spin dynamics is then studied
by measuring the decay of $\rho_{c}$ for the $\sigma^+$-polarized
probe pulse delayed in time relative to the pump pulse. Pump
(probe) power is kept at $1$ ($0.05$)~mW, for which $B_{N} \approx
0$ [Fig.~\ref{dnp}(inset)]. Also, the probe power being $20$ times
smaller than the pump power, it does not destroy the pump-induced
spin polarization. A schematic of the pump and probe pulse
configurations is shown in the inset of Fig.~\ref{trpl}.
Exploiting this method we study the spin dynamics in a wide
time-range from picoseconds to
milliseconds.~\cite{ikezawaprb72,paljpsj75}

For the study of spin dynamics in the nanosecond time-regime, pump
and probe pulses are derived from a picosecond Ti:Sapphire laser
and the pump-probe delay $\tau$ is controlled by optical delay.
The polarization-selected PL originating from the probe pulse is
time-resolved in a streak camera to monitor the kinetics of
$\rho_{c}$ for the probe-generated PL. Figure~\ref{trpl} shows
such kinetics of $\rho_{c}$ at $\tau = 2$~ns for
$\sigma^{+}$-polarized probe pulse when the pump pulse is
co-circularly ($\sigma^{+}$) or cross-circularly ($\sigma^{-}$)
polarized. Data are taken for $B_{\text{ext}} = 0$ and $0.1$~T. As
seen in Fig.~\ref{trpl}, at times beyond $300$~ps, kinetics of
$\rho_{c}$ reaches a constant value (refer to as $A_{\text{CP}}$
in Fig.~\ref{plspec}), which is strongly negative (positive) for
the co- (cross-) circularly polarized pump-probe excitation. The
difference $\Delta A_{\text{CP}}$ between the $A_{\text{CP}}$ for
the two cases (co- and cross-circularly polarized pump-probe
configurations) can be used as a measure of the pump-induced spin
polarization of the resident electrons. We find that for a given
$\tau$, $\Delta A_{\text{CP}}$ is sensitive to $B_{\text{ext}}$.
The difference $\Delta A_{\text{CP}}$(0~T) measured at
$B_{\text{ext}} = 0$~T is approximately one-third of $\Delta
A_{\text{CP}}$(0.1~T) measured at $B_{\text{ext}} = 0.1$~T for
$\tau = 2$~ns [Fig.~\ref{trpl}]. An identical behavior is seen
when measured at $\tau = 10$~ns. Following the theory of Merkulov
{\etal}~\cite{merkulovprb65} we conclude that for $B_{\text{ext}}
= 0$, a {\emph{partial relaxation}} (up to one-third of the
initial value) of electron spin polarization due to the FFNS takes
place within a time shorter than $2$~ns. However, the remaining
spin polarization does not decay up to $10$~ns. In presence of
$B_{\text{ext}} = 0.1$~T, electron spin relaxation by the FFNS is
suppressed. In that case electron spin relaxation time is much
longer than $10$~ns.

As discussed in Sec.~\ref{intro}, for $B_{\text{ext}} = 0$ decay
of the electron spin polarization surviving after $2$~ns
(one-third of the initial value) is governed by the NSPE over a
longer timescale.~\cite{merkulovprb65} For the study of spin
dynamics in the microsecond timescale, it is more convenient to
measure the magnetic field- and delay-dependence of $\rho_{c}$
integrated over the PL lifetime. In these experiments, the pump
and probe pulses are derived from a CW Ti:Sapphire laser by using
acousto-optic modulators (AOMs) which act as electrically
controlled gates. Pump and probe pulse widths ($1$~$\mu$s each)
and the delay $\tau$ between them are controlled by sending
electrical pulses to the AOMs from a programable function
generator. In this case the accessible range of $\tau$ is not
limited by the laser pulse repetition period of $12$~ns. Details
of the experimental setup are discussed in
Ref.~\onlinecite{paljpsj75}. We measure the difference $\Delta
\rho_{c}$ between $\rho_{c}$ (integrated over the PL lifetime) for
the probe PL for cross- and co-circularly polarized pump-probe
excitations.

\begin{figure}[tbh]
\includegraphics[clip,width=8.2cm]{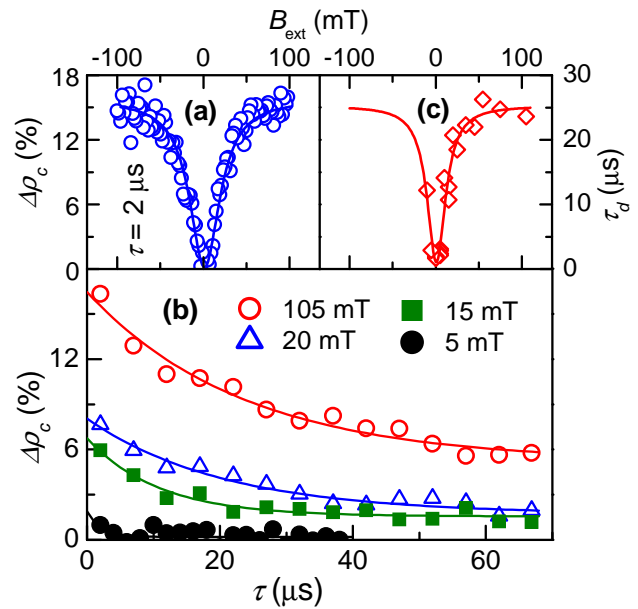}
\caption{\label{taudep}(Color online) (a) Dependence of $\Delta
\rho_{c}$ on $B_{\text{ext}}$ for $\tau = 2$~$\mu$s. The solid
line is a Lorentzian fit. Total depolarization ($\Delta \rho_{c}
\approx 0$) of the electron spin is seen at $B_{\text{ext}} = 0$.
(b) Dependence of $\Delta \rho_{c}$ on $\tau$ at a few values of
$B_{\text{ext}}$. Solid lines are exponential fits characterized
by the spin decay time $\tau_{d}$, which is plotted in (c) as a
function of $B_{\text{ext}}$. The solid line in (c) is a
Lorentzian fit.}
\end{figure}

At first we measure the dependence of $\Delta \rho_{c}$ on
$B_{\text{ext}}$ at $\tau = 2$~$\mu$s [Fig.~\ref{taudep}(a)]. Near
$B_{\text{ext}}=0$ we find that $\Delta \rho_{c} \approx 0$. This
suggests that for $B_{\text{ext}} = 0$, a {\emph{total
depolarization}} of electron spins takes place within $2$~$\mu$s.
This is caused by the NSPE, indicating that the Larmor precession
period $T_{N}$ of nuclear spins in the hyperfine field of
electrons is shorter than $2$~$\mu$s.

To obtain a more quantitative estimate of $T_{N}$, we perform a
systematic measurement of electron spin polarization decay time
$\tau_{d}$ as a function of $B_{\text{ext}}$.
Figure~\ref{taudep}(b) shows the delay dependence of $\Delta
\rho_{c}$ at different values of $B_{\text{ext}}$. The decay of
$\Delta \rho_{c}$ with $\tau$ can be well approximated by an
exponential function, $\Delta \rho_{c} = A_{0} + A_{1}
\exp(-\tau/\tau_d)$ [Fig.~\ref{taudep}(b)]. The decay time
$\tau_{d}$ obtained from such fits is shown in
Fig.~\ref{taudep}(c) for different values of $B_{\text{ext}}$. It
is seen that $\tau_{d}$ decreases down to $1$~$\mu$s when $\left|
B_{\text{ext}} \right|$ goes to zero, for which electron spin
relaxation takes place due to hyperfine interaction with
nuclei.~\cite{highfield} At low temperatures, the {\emph{total
depolarization}} of the electron spins under the vanishing
external magnetic field, is caused by the NSPE in the microsecond
timescale.~\cite{merkulovprb65} Therefore, we assign the time
$1$~$\mu$s obtained from Fig.~\ref{taudep}(c) at $B_{\text{ext}} =
0$ as an upper limit of $T_{N}$.~\cite{accuracy} No other
experimental study of $T_{N}$ in QDs is available to our knowledge
in the literature for a comparison with our estimate of $T_{N}$.
However, the value of $T_{N} \sim 1$~$\mu$s obtained by us for the
self-assembled InP QDs is comparable to the theoretical estimate
of $T_{N}$ for GaAs QDs.~\cite{merkulovprb65} This agreement
suggests that $T_{N}$ may have the same order of magnitude ($\sim
1$~$\mu$s) in different self-assembled III-V quantum dots.

\section{Conclusion}
The effects of nuclear spins on the electron spin dynamics in
singly negatively charged InP QDs are studied at low temperature
($T \approx 5$~K), where hyperfine interaction with the nuclear
spins is the dominant relaxation channel for the electron spins.
We observe that at the vanishing external magnetic field, partial
relaxation (up to one-third of the initial value) of electron spin
polarization takes place within $\sim 1$~ns due to the frozen
fluctuations of the nuclear hyperfine field. A value of $15$~mT is
estimated for the effective magnetic field $B_{f}$ of the frozen
fluctuations of nuclear spins. Total depolarization of electron
spins with a characteristic time of $1$~$\mu$s is observed at the
vanishing external magnetic field, due to the slow variation of
$B_{f}$ in time caused by the nuclear spin precession in the
hyperfine field of electrons. The characteristic  time of
$1$~$\mu$s is assigned to the nuclear spin precession period
$T_{N}$ in the hyperfine field of electrons. At high excitation
power the dynamic nuclear polarization is observed, giving rise to
an Overhauser field $B_{N} = 6$~mT at $50$~mW excitation.

\begin{acknowledgments}
Authors thank V. K. Kalevich and I. Ya. Gerlovin for fruitful
discussions. The work is partially supported by ``Grant-in-Aid for
Scientific Research'' Nos.~\mbox{17-5056}, 13852003, and 18204028
from the MEXT of Japan and ``R{\&}D Promotion Scheme Funding
International Joint Research'' promoted by NICT of Japan, by ISTC,
grant 2679, by Russian Ministry of Sci.~{\&}~Edu., grant
RNP.2.1.1.362, and by RFBR, grant 06-02-17137-a.
\end{acknowledgments}


\begin{thebibliography}{99}
\bibitem{optpump} This behavior is similar to that observed
for optical pumping of electron spins in $n$-type
semiconductors.~\cite{optorientbook} We estimate that the number
of photocreated electron-hole pair per QD per pulse remains less
than one up to the highest average excitation power (50~mW) used.
\bibitem{altexct} The minimum of $\left| \rho_{c} \right|$ appears
at $B_{\text{ext}} \approx 0$ when the polarization of the
excitation beam is modulated between $\sigma^{+}$ and $\sigma^{-}$
at a frequency of $42$~kHz, by using a photo-elastic modulator in
the excitation path. It is well known that in such a case dynamic
nuclear polarization cannot build up.~\cite{optorientbook}
\bibitem{highfield} For $B_{\text{ext}} \ne 0$, the effect of
hyperfine interaction on the electron spin relaxation begins to
diminish, because the external magnetic field starts to influence
the electron-nuclear spin dynamics. Electron spin decay time
$\tau_{d}$ measured for $\left| B_{\text{ext}} \right| > 0$ cannot
be directly assigned to the nuclear spin precession period $T_{N}$
in the hyperfine field of electrons. Effects of the external
magnetic field on the electron-nuclear spin dynamics is discussed
in Ref.~\onlinecite{merkulovprb65}.
\bibitem{accuracy} Accuracy of this estimate is decided by the
accuracy of the exponential fit, estimated at $30${\%} in
Fig.~\ref{taudep}(b).
\end{thebibliography}
\end{document}